\documentclass[num-refs]{wiley-article}

\usepackage{graphicx}
\usepackage{amsmath}

\usepackage{natbib}
\usepackage{url}
\usepackage{cite}
\usepackage{changes}
\usepackage{amssymb}
\usepackage{subfig}
\usepackage{etoolbox}
\usepackage{comment}
\usepackage{siunitx}
\usepackage{verbatim,color,amssymb}
\usepackage{graphicx}
\usepackage{epstopdf}
\usepackage{epsfig}
\usepackage{subfig}
\usepackage{mathtools}
\usepackage{amssymb,amsmath}
\usepackage{textcomp}
\usepackage{amsmath}
\usepackage{rotating}
\usepackage{array}
\usepackage{multirow} 

\newcommand\BibTeX{{\rmfamily B\kern-.05em \textsc{i\kern-.025em b}\kern-.08em
		T\kern-.1667em\lower.7ex\hbox{E}\kern-.125emX}}

\papertype{Original Article}

\title{Clustering functional data with measurement errors: a simulation-based approach}

\author[1]{Tingyu Zhu}
\author[2]{Lan Xue$^*$}
\author[3]{Carmen Tekwe}
\author[4]{Keith Diaz}
\author[5]{Mark Benden}
\author[6]{Roger Zoh}

\affil[1]{Department of Statistics, Oregon State University, Oregon, USA}

\affil[2]{Department of Statistics, Oregon State University, Oregon, USA}

\affil[3]{Department of Epidemiology and Biostatistics, Indiana University, Indiana, USA}

\affil[4]{Department of Medicine, Columbia University Medical Center, New York}

\affil[5]{ Department of Environmental \& Occupation Health, Texas A\&M University, Texas, USA}

\affil[6]{Department of Epidemiology and Biostatistics, Indiana University, Indiana, USA}

\corraddress{Lan Xue, Department of Statistics, Oregon State University, Corvallis, OR 97330, USA.} \corremail{xuel@stat.oregonstate.edu}

\fundinginfo{NSF, DMS, Award Number: 1613190, 1821198, 1812258; National Heart, Lung and Blood Institute, Award Number: \#NO1-HC-55023, \#U01-HL48941-44. }

\runningauthor{Xue et al.}

\begin{document}

\maketitle

\begin{abstract}
Clustering analysis of functional data, which comprises observations that evolve continuously over time or space, has gained increasing attention across various scientific disciplines. Practical applications often involve functional data that is contaminated with measurement errors arising from imprecise instruments, sampling errors, or other sources. These errors can significantly distort the inherent data structure, resulting in erroneous clustering outcomes. In this paper, we propose a simulation-based approach designed to mitigate the impact of measurement errors. Our proposed method estimates the distribution of functional measurement errors through repeated measurements. Subsequently, the clustering algorithm is applied to simulated data generated from the conditional distribution of the unobserved true functional data given the observed contaminated functional data, accounting for the adjustments made to rectify measurement errors. Our simulation studies show that the proposed method has improved numerical performance than the naive methods that neglect such errors. Our proposed method was applied to a childhood obesity study, giving more reliable clustering results. 

\end{abstract}

\section{INTRODUCTION}\label{section:intro}

In recent years, the field of biomedical research has experienced a surge in the accessibility of complex function-valued data, largely driven by technological advancements. In particular, the widespread adoption of wearable devices has revolutionized various biomedical studies aimed at monitoring individuals' health conditions. By facilitating the collection of detailed and continuous data, these wearable devices provide researchers with valuable insights into diverse aspects of health and well-being.

Despite the potential of wearable devices, the analysis of functional data derived from these measurements presents a significant challenge. One major obstacle is due to the inherent measurement errors associated with these devices. It is widely acknowledged that the physical activity (PA) measurements captured by wearable devices serve as mere approximations of the true PA of the individual \citep{bassett2012, crouter2006estimating, jacobi2007physical,warolin2012,rothney2008validity}. As a result, it becomes imperative to consider and address these measurement errors to ensure the accuracy and reliability of the data analysis.

There has been a recent advance in literature to address measurement errors within functional data analysis. The majority of research efforts have been dedicated to integrating measurement error considerations into regression problems, encompassing a wide range of applications. In particular, \citet{tekwe2018functional} proposed a likelihood-based estimation method to fit functional multiple indicators, multiple causes measurement error (FMIMIC ME) model using the EM algorithm. \citet{tekwe2019instrumental} corrected the attenuation bias in a scalar-on-function regression
model using a method of moment approach, while \citet{jadhav2022function} developed a two-step estimation based on regression calibration. In addition, \citet{JadhavMa20} extended the conditional-score method to estimate the generalized functional linear models in the presence of measurement error, and  \citet{Zhangpl2023} proposed a corrected score approach for the partially functional linear quantile regression model with measurement errors.

Our research is driven by a childhood obesity study, with the primary objective of exploring the connection between daily energy expenditure (DEE) and the subsequent development of obesity. Growing evidence underscores the importance of personalized DEE recommendations for effective obesity prevention strategies. To facilitate these tailored recommendations, it becomes essential to accurately identify subgroups displaying distinct latent patterns of DEE. This identification is critical as it provides deeper insights into the DEE-obesity relationship, enabling us to customize interventions accordingly. As DEE cannot be directly observed, the children were provided with accelerometer armbands to approximate their DEE. Our project aims to develop statistical methods to cluster individuals based on DEE data acquired from wearable devices while considering the data as functional data susceptible to measurement errors. Our additional goal is to mitigate the impact of measurement errors in DEE to improve the accuracy of obesity prediction.

Clustering functional data has been an active area of research in recent years. Functional data, where observations are represented as functions rather than traditional data vectors, present unique challenges for clustering due to their infinite-dimensional nature.   \citet{Abraham2003, Luan2003, Coffey2014} propose to first reduce the dimensionality of the function data through discretization or projecting each curve onto a finite-dimensional basis. Subsequently, conventional clustering methods for multivariate data are performed with the resulting discretized data or basis coefficients.  \citet{Fraley2002}  proposed a model-based clustering approach for functional data using Gaussian mixtures to identify clusters based on underlying probability density functions.   \citet{JamesSugar2003} presented a clustering method based on random effects models for sparse and irregularly observed functional data. Meanwhile, \citet{DelaigleHall2012} offered theoretical insights into the performance of the classification method, providing mathematical justifications or statistical guarantees. 

The aforementioned clustering methods require the specification of the number of clusters in advance, which can be challenging in cluster analysis. More recently,  \citet{zhuqu2018}  introduced a novel regression-based approach to cluster functional data by a  pairwise penalization method. It approximates functional data with polynomial splines and partition observations into subgroups by penalizing the pairwise distances between the B-spline coefficient vectors. One advantage of this approach is its ability to automatically determine the number of clusters, eliminating the need for pre-specification and enabling simultaneous model estimation and subject subgrouping. Moreover, the proposed method demonstrates its versatility by characterizing longitudinal trajectories, even when dealing with unbalanced data.

However, when the functional data are subject to measurement errors, the aforementioned methods can not be applied directly. The presence of measurement errors can introduce biases and distortions in the data, which, in turn, can misrepresent the true underlying patterns and structures. Moreover, measurement errors can lead to increased variability within clusters, making it difficult to distinguish underlying subgroups. Consequently, the overall clustering quality may be adversely impacted, leading to less distinct and less informative clusters.

This paper aims to introduce a simulation-based approach for clustering functional data that involves measurement errors. We first estimate the distribution of the measurement errors based on repeated measurements, which enables us to simulate additional functional data. Simulated pseudo-functional data has reduced variability compared to the original measured functional data, resulting in improved clustering outcomes. Although our primary focus lies on the pair-wise penalization method presented in \citet{zhuqu2018}, the proposed method can be readily adapted to various other clustering techniques to effectively incorporate measurement errors. 

A similar approach was used in \citet{SuCarroll2018} to cluster multivariate vector data with complex measurement errors. Our method can be viewed as an extension of that method to deal with functional data. Furthermore, we provide an in-depth exposition on the estimation of measurement error distributions based on repeated measurements of the functional surrogate. Estimating the error distributions presents a significant challenge due to the inherent data heterogeneity stemming from the presence of subgroups while the labels of the subgroups are unknown to us. To address this challenge, we introduce an interactive procedure for estimating error distributions and conducting subsequent clustering analysis.


The rest of the paper is organized as follows. The model setup is introduced in Section 2. The methodology for clustering functional data with measurement error is presented in Section 3. Section 4 presents results from simulation studies, while Section 5 applies the proposed method to a childhood obesity study. We provide some concluding remarks and discussions in Section 6.

\section{THE MODEL}
For subject $i$ ($i=1,\cdots,n$), let $Y_{i}=\left\{Y_{i}(t),t\in \mathcal{I}\right \}$ be a continuous random function observed on a bounded closed interval $\mathcal{I} $ in $\mathcal{R}$. Without loss of generality, let $\mathcal{I}=[0,1]$. We assume a subject-specific model for the functional data. In particular, for subject $i$ and $t\in[0,1]$,
\begin{equation*}
    Y_{i}(t)=f_{i}(t)+\epsilon_i(t),
\end{equation*}
where $f_{i}(t)$ is a subject-specific smooth mean function, and the error term $\epsilon_i(t)$ is assumed to be a Gaussian process with zero mean and variance function $\Sigma(t,s)$. 

We assume that the $n$ subjects form $K$ distinct groups, where the number of groups $K$ and the partition of these $K$ groups are both unknown. The subjects in the same group share the same mean trajectory. That is, the subjects $i,j$ belong to the same group if and only if $f_{i}(t)=f_{j}(t)$ for all $t\in [0,1]$. Let $f_1,\cdots, f_K$ be the group means of these $K$ distinct groups. 

In this paper, we deal with a more challenging situation where the functional data $Y_i$ is unobservable; instead, a functional surrogate is observed, which is prone to additive measurement errors. Let
\begin{equation*}
    W_{i}(t)=Y_{i}(t)+U_i(t),
\end{equation*}
where $U_i(t)$ is a functional measurement error term. We assume $U_i(t)$ is independent of $Y_i(t)$ and follows a Gaussian process with zero mean and variance function $\Sigma_U(t,s)$.  We allow the measurement errors to be correlated over time and do not impose any structure on the covariance function $\Sigma_U(t,s)$.

Our goal is to perform cluster analysis based on error-prone measures  $\left\{W_{i}(t)\right\}_{i=1}^{n}$. We propose a simulation-based procedure for cluster analysis. It simulates synthetic functional data from a distribution that adjusts the measurement error. In addition, the proposed clustering method does not require the specification of the number of clusters in advance and allows simultaneous estimation of the number of clusters and the group mean trajectories.

\section{METHODS}
When there is no measurement error, \citet{zhuqu2018} proposed a pairwise-grouping penalization method for clustering longitudinal profiles with subgroups. In this paper, we extend it to accommodate measurement errors. The method starts with approximating the subject-specific mean trajectory using polynomial splines. The spline functions provide effective approximations to smooth functions. Let $B_1,\ldots, B_{J_n}$ be a set of B-spline basis. Then
\[
f_i(t) \approx g_{i}(t)=\sum_{l}^{J_{n}}B_{l}(t)\beta_{il}={\bf B}^{T}(t){\bf \beta}_{i},
\]
for a set of subject-specific B-spline coefficients ${\bf \beta}_{i}=\left(\beta_{i1},\cdots,\beta_{iJ_n}\right)^{T}$, and ${\bf B}(t)=\left(B_1(t),\cdots,B_{J_{n}}(t)\right)^{T}$. Then, individuals in the same group share similar B-spline coefficients. The pairwise-grouping penalization method by Zhu and Qu (2018)  performs cluster analysis by minimizing the following objective function,
\begin{equation}
    L({\bf\beta})=\frac{1}{2}\sum_{i=1}^{n}\left\{\|{\bf y}_i-{\bf B}_i\beta_i\|_2^2+\lambda_1\beta_i^{T}D_{d}\beta_{i}\right\}+\sum_{i,j\in \mathcal{L}}\rho(\beta_i-\beta_j,\lambda_2),\label{equ:pgp}
\end{equation}
where ${\bf y}_i=\left(Y_{i}(t_{i1}),\cdots,Y_{i}(t_{in_{i}})\right)^{T}$, ${\bf B}_i=\left(B(t_{i1}),\cdots,B(t_{in_{i}})\right)^{T}$, and $D_{d}=\Delta_{d}^{T}\Delta_{d}$ with $\Delta_{d}$ being the $d$th-order difference operator. In addition, $\rho(\cdot)$ is a penalty function and depends on a tuning parameter $\lambda_2$, and $\mathcal{L}=\left\{(i,j), 1\leq i<j\leq n\right\}$ is the index set containing all distinct pairs of $n$ individuals. The second term on the right-hand side of (\ref{equ:pgp}) penalizes excessive roughness or complexity of the curves and improves spline fitting performances by preventing overfitting. The tuning parameter $\lambda_1$ controls the balance between the goodness of fit and the complexity of the model. The final term in (\ref{equ:pgp}) applies a penalty to the pairwise differences between B-spline coefficients, promoting identical B-spline coefficients among different individuals, thus encouraging the formation of clusters. Similar to \citet{zhuqu2018}, we consider the minimax concave penalty for $\rho(\cdot)$, which has sparsity property and is nearly unbiased. 

When encountering measurement errors in the functional data, obtaining additional information for identifying the measurement error becomes necessary. This additional information can be in the form of instrument variables or repeated measures \citep{carroll2006measurement, tekwe2019instrumental, Zhangpl2023}. This paper focuses on the scenario where repeated measures are accessible for each functional data set. That is, for each $i=1,\cdots, n$, there are $J$ repeated measures  
\begin{equation*}
    W_{ij}(t)=Y_{i}(t)+U_{ij}(t),
\end{equation*}
where $j=1,\cdots, J$, and the functional measurement errors $\left\{U_{ij}(t), \ t\in [0,1]\right\}_{i,j=1}^{n,J}$ are independent and follow Gaussian process with zero mean and covariance function $\Sigma_{U}(t,s)$.

Let $\bar{W}_i(t)=\sum_{j=1}^{J}W_{ij}(t)/J$ be the averaged functional data over $J$ repeated measures. Naive clustering procedures can be considered based on either the averaged functional data $\left\{\bar{W}_i(t)\right\}_{i=1}^{n}$ or a single measure with error $\left\{W_{i1}(t)\right\}_{i=1}^{n}$. In particular, let $\hat{\mathcal{G}}_{Ave}$ be a partition of $\left\{1,\cdots, n\right\}$ and represent the clustering result of applying the pairwise grouping penalized method to the averaged functional data, which replaces ${\bf y}_i$ with $\bar{W}_i$ in (\ref{equ:pgp}). Similarly, define $\hat{\mathcal{G}}_{W_1}$ be the clustering result based on a single measure $W_{i1}(t)$. Both naive clustering procedures ignore the measurement error in functional data. We propose a simulation-based approach to correct the functional measurement errors, ultimately enhancing clustering performance. 

Let $\hat{\mathcal{G}}^{0}$ be an initial clustering result. For example, $\hat{\mathcal{G}}^{0}=\hat{\mathcal{G}}_{Ave}$. Denote $\hat{\mathcal{G}}^{0}=\left\{\hat{{g}}_1^{0},\cdots, \hat{{g}}_{\hat{K}}^{0}\right\}$, a partition of $n$ individuals with $\hat{K}$ subgroups. The initial clustering result can be used to estimate the distribution of the functional data $Y_i(t)$. In particular, let $\bar{{W}}^k(t)$ be the group mean of the $k$th cluster with 
\[\bar{{W}}^k(t) = \frac{1}{|\hat{g}_k^{0}|} \sum_{i \in \hat{g}_k^{0} } \bar{W}_i(t),\]
where $|\hat{g}_k^{0}|$ denotes the number of elements in $\hat{g}_k^{0}$.

In addition, repeated measures can be used to estimate the covariance structure of measurement errors. In particular, we use the within-subject sample covariance matrix with $\hat{\Sigma}_U(t,s) = \frac{1}{n(J-1)}\sum_{i=1}^n \sum_{j=1}^J \left({W}_{ij}(t)-\overline{{W}}_i^*(t)\right)\left({W}_{ij}(s)-\overline{{W}}_i^*(s)\right),$ where $\overline{{W}}_i^*(t)=\frac{1}{J} \sum_{j=1}^J{W}_{ij}(t)$. When there are multiple subgroups, the data is inhomogeneous with different group means. Therefore, the usual between-subject sample covariance is no longer valid for estimating the variance of $\epsilon(t)$. However, based on the initial clustering results, the individuals in the same subgroup are of similar group means. Therefore, for the $j$th repeated measure and the $k$th subgroup in $\hat{\mathcal{G}}^{0}$, we consider
$\hat{\Sigma}^{jk}_W(t,s)=\frac{1}{(|\hat{g}^{0}_k|-1)}\sum_{i \in \hat{g}^{0}_k} \left({W}_{i j}(t)-\bar{{W}}_j^k(t)\right)\left({W}_{i j}(s)-\bar{{W}}_j^k(s)\right)$ with $\bar{{W}}_j^k(t) = \frac{1}{|\hat{g}^{0}_k|} \sum_{i \in \hat{g}^{0}_k} {W}_{ij}(t)$, which is the sample variance of $W_j(t)$ in the $k$th subgroup. Note that $\hat{\Sigma}^{jk}_W(t,s)$ has a reasonable performance when the group size $|\hat{g}^{0}_k|$ is at least three. Thus, to estimate the covariance matrix of $W$, we consider averaging across all repeated measures and all subgroups with a size of at least three. That is, $\hat{\Sigma}_W (t,s) = \frac{1}{J|\mathcal{B}^{0}|}\sum_{k \in \mathcal{B}^{0}} \sum_{j=1}^J \hat{\Sigma}^{jk}_W(t,s),$ where $\mathcal{B}^{0}$ is the index set of subgroups with a size of at least three.  Consequently, we have $\hat{\Sigma}_{\epsilon}(t,s) = \hat{\Sigma}_W(t,s)-\hat{\Sigma}_U(t,s)$.

In particular, if individual $i$ belongs to the $k$th group for some $k=1,\cdots, \hat{K}_{n}$, then 
${Y}_i(t) \sim MVN(\bar{{W}}^k(t), \hat{\Sigma}_{\epsilon})$ approximately, where $\bar{{W}}^k(t)$ is the group mean that the $i$th subject belongs to. Therefore, we propose to simulate pseudo functional data $\left\{{Y}^*_i(t)\right\}_{i=1}^{n}$ from the estimated 
  conditional distribution  ${Y}_{i}(t)$ given $\bar{{W}}_i$, which is $MVN(\bar{\mu}_i(t), \Bar{\Sigma})$ with
 $\bar{\mu}_i(t) = \bar{{W}}^k(t)+ \hat{\Sigma}_{\epsilon} (\hat{\Sigma}_{\epsilon} +  \hat{\Sigma}_U/J)^{-1}(\bar{{W}}_i(t) -\bar{{W}}^k(t) )$ and $\Bar{\Sigma} = \hat{\Sigma}_{\epsilon} - \hat{\Sigma}_{\epsilon}  (\hat{\Sigma}_{\epsilon} +  \hat{\Sigma}_U/J)^{-1}\hat{\Sigma}_{\epsilon} $.

Furthermore, when the size of the $k$th group is 1, it means this subgroup consists of a sole individual, denoted as $i$. We propose to pool it towards the closest group that has more than two individuals by replacing the original group mean $\bar{{W}}^k(t)$ with a pooled group mean, defined as $\bar{{W}}_p^k(t)$. Then, $\bar{\mu}_i = \bar{{W}}_p^k(t)+ \hat{\Sigma}_{\epsilon} (\hat{\Sigma}_{\epsilon} +  \hat{\Sigma}_U/J)^{-1}(\bar{{W}}_i(t) -\bar{{W}}_p^k(t) )$. Here, the pooled mean $\bar{{W}}_p^k(t)$ for group $k$ is calculated as $\bar{{W}}_p^k(t) = w_i\bar{{W}}_{i}(t) + (1-w_i)\bar{{W}}^{k^*}(t)$, where $\bar{{W}}^{k^*}(t)$ is the average of the group that is closest to individual $i$, and $0\leq w_i\leq 1$ is a weight used when calculating the pooled average.

\subsection{Algorithm}
Here is the summary of the algorithm for our proposed method. 
\begin{enumerate}
\item Conduct initial clustering and obtain initial clustering result $\hat{\mathcal{G}}^{0}=\left\{\hat{{g}}_1^{0},\cdots, \hat{{g}}_{\hat{K}}^{0}\right\}$. 
For example, it can be based on 
 single measures $\left\{{W}_{i1}(t)\right\}_{i=1}^{n}$ or on the average values from $J$ replicated measures, defined as $\left\{\bar{{W}}_{i}(t) = \frac{1}{J} \sum_{j=1}^J {W}_{ij}(t)\right\}_{i=1}^{n}$. 

\item Obtain the estimated covariance matrices of $\Sigma_U$ and $\Sigma_W$ based on the initial clustering results in Step 1. 

\item Simulate  ${Y}^*_i(t)$ from the estimated conditional distribution  ${Y}_{i}(t)|\bar{{W}}_i(t)$ or ${Y}_{i}(t)|{{W}}_{i1}(t)$, for $i=1,\ldots,n.$ 

\item Conduct cluster analysis on $\left\{{Y}^*_i(t)\right\}_{i=1}^{n}$. 

\item Repeat Steps 2-4 until 

(1) the membership of each subject in the $r$th iteration and the $(r-1)$th iteration are the same; or (2) the BIC in the $r$th iteration is higher than the BIC in the $(r-1)$th iteration.
\end{enumerate}

\subsection{Parameter selection}

The effectiveness of the pairwise-grouping penalization method crucially depends on the selection of two tuning parameters, namely $\lambda_1$ and $\lambda_2$ in (\ref{equ:pgp}). However, conducting a grid search for both parameters simultaneously can significantly escalate computational expenses. Following \citet{zhuqu2018}, we employ a two-step procedure. In the first step, we focus on finding the optimal value of $\lambda_1$ with $\lambda_2$ set to 0, accomplished by minimizing,
$$B I C\left({\lambda_1}\right)=\sum_{i=1}^n\left\{\log \left(\frac{\left\|\mathbf{y}_i-\hat{\mathbf{f}}_i\right\|_2^2}{n_i}\right)+\frac{\log \left(n_i\right)}{n_i} \mathrm{df}_i\right\},$$
where $\left\|\mathbf{y}_i-\hat{\mathbf{f}}_i\right\|_2^2=\sum_{j=1}^{n_i}\left(Y_{i}(t_{ij})-\hat{f}_{i}(t_{ij})\right)^2$ with  $n_i$ being the number of time points observed for individual $i$ and $\operatorname{df}_i=\operatorname{tr}\left\{\mathbf{B}_i\left(\mathbf{B}_i^T \mathbf{B}_i+\lambda_1 D_d\right)^{-1} \mathbf{B}_i^T\right\}$.

Then, select $\lambda_2$ with the optimal $\lambda_1$ from the first step by minimizing 
$$B I C \left({\lambda_2}\right)=\log \left(\frac{\sum_{i=1}^{n}\|\mathbf{y}_i-\hat{\mathbf{f}}_i\|_2^2}{N}\right)+\frac{\log (N)  \mathrm{df}}{N},$$
where  $N=\sum_{i=1}^{n}n_i$ is the total of number of observations, and $\mathrm{df}=\frac{\hat{K}}{n} \sum_{i=1}^n \mathrm{df}_i$ with $\hat{K}$ being the number of groups identified. 


\section{SIMULATION STUDIES}

In this simulation, functional data are simulated from two groups, each containing 20 or 40 subjects, which results in a sample size of $n=40$ or $80$. The two groups are characterized by two distinct mean functions $f_{(1)}(t) = \cos(1.5 \pi t) + 2.5$ and $f_{(2)}(t) = \sin(1.5 \pi t) + c$. To investigate how the distance between two groups affects the performance of the proposed method, we consider two different values for $c$, where $c=-0.5$ represents the "close" case and $c=-3.5$  for the "far" case.  For observations in the $k$-th ($k=1,2$) group, the true functional response $y_{i}(t_{m})$ for subject $i$ at time $t_m$  is generated by $y_{i}(t_{m})=f_{(k)}\left(t_{m}\right)+\varepsilon_{i m}$ with $ m=1, \cdots, 10$, where random errors within subjects are independent $\varepsilon_{i m} \sim^{i i d} N\left(0,0.4^2\right)$, and $\left\{t_{ m}\right\}_{m=1}^{10}$ are equally spaced points on $[0,1]$. 

In addition, for each subject $i$, three repeated measures contaminated with measurement error are generated by $w_{ij}(t_m) = y_{i}(t_m) + u_{ij}(t_m)$ for $j=1,2,3$, where the measurement errors $u_{ij}(t_m)$ are of zero means and a compound symmetry correlation structure with $\rho=0.5$ over time. In particular, $u_{ij}(t_m)=\gamma_{ij}+ \gamma^{'}_{ij}(t_m) $ with $\gamma_{ij}\sim^{i i d} N\left(0,\sigma^2/4\right)$ and $\gamma^{'}_{ij}(t_m) \sim^{i i d} N\left(0,\sigma^2/4\right)$.  Here $\gamma_{ij}$ is a common random variable shared across different time points, therefore leading to a compound symmetry correlation structure. Simulations are conducted for $\sigma= 2$ and $3$, resulting in variance ratios of $12.5$ and $28.1$ between the measurement error and the true functional data, respectively.


We compare the clustering results from several different methods. The oracle method (ORACLE) applies the pairwise grouping penalized method to true functional data $\left\{{Y}_i(t)\right\}_{i=1}^{n}$, which is not available in practice and only used for comparison purposes. We also consider two naive methods that apply the pairwise grouping penalized method to the single observed data ${{W}}_{i1}(t)$, or the average data over three replicates $\bar{{{W}}}_{i}(t)$. Both naive methods ignore measurement errors and serve as the baseline to assess the effectiveness of the proposed method. For our proposed method, we consider pseudo observations $\left\{{Y}^*_i(t)\right\}_{i=1}^{n}$ 
 simulated from the estimated conditional distribution of ${Y}_i(t)|{{W}}_{i1}(t)$ or ${Y}_i(t)|\bar{{W}}_{i}(t)$.

To evaluate the accuracy of different clustering methods, we calculate clustering indexes that are frequently used in the literature.  In particular, for a given clustering membership of subjects, define true positive (TP), true negative (TN), false positive (FP), and false negative (FN) as follows, where TP is the number of pairs that are correctly clustered together;  TN is the number of pairs of subjects that are correctly not clustered together, that is those are from different clusters and assigned to different clusters; FP is the number of pairs of subjects from different clusters but assigned to the same cluster; FN is the number of pairs of subjects from the same cluster but assigned to different clusters. We use the following indexes to evaluate the clustering performance with the Rand index \citep{Rand1971} $Rand =\frac{TP+TN}{TP+TN+FP+FN}$; adjusted Rand index  \citep{HubertArabie1985} $aRand =\frac{Rand - E(Rand)}{\max(Rand) -E(Rand)}$; and Jaccard index \citep{Jaccard1912} $Jaccard = \frac{TP}{TP+FP+FN}$. For all three indexes, a higher value indicates a better agreement between the estimated and the true group memberships. The simulation results are based on 50 replications.

\begin{table}
\centering
\begin{tabular}{|r|r|r|r|cccc|}
\hline
Model &$\sigma$ & n & Method & \# clusters & Rand & Jaccard & aRand \\
   \hline
\multirow{12}{*}{Far}& \multirow{6}{*}{2}& \multirow{3}{*}{40} & ${Y}_i(t)$  &  2 & 1  &  1 & 1\\
 & & &$\bar{{W}}_{i}(t)$& 4.48 & 0.924 & 0.844 & 0.846\\
 & & &${Y}^{*}_i(t)$ & 2.42 & 0.972 & 0.943 & 0.944 \\
 \cline{3-8}
 &    & \multirow{3}{*}{80} &  ${Y}_i(t)$  &  2 &1   &1   & 1\\
   &   &  & ${\mathbf{W}}_{i}(t)$ & 6.34 & 0.928 & 0.855 & 0.856\\
  &   &  & ${Y}^{*}_i(t)$ & 2.46 & 0.970 & 0.938 & 0.939\\
\cline{2-8}
& \multirow{6}{*}{3}& \multirow{3}{*}{40} & $\mathbf{Y}_i(t)$  & 2  &  1 &  1 & 1\\
 & & &$\bar{{W}}_{i}(t)$& 5.36 & 0.918 & 0.833 & 0.836\\
 & & &${Y}^{*}_i(t)$ & 2.74 & 0.978 & 0.956 & 0.957 \\
 \cline{3-8}
 &    & \multirow{3}{*}{80} & ${Y}_i(t)$  & 2  &  1 & 1  & 1\\
   &   &  & $\bar{{W}}_{i}(t)$ & 7.64 & 0.917 & 0.833 & 0.834\\
  &   &  & ${Y}^{*}_i(t)$ & 2.54 & 0.975 & 0.949 & 0.950 \\
   \hline
    \hline
\multirow{12}{*}{Close}& \multirow{6}{*}{2} & \multirow{3}{*}{40} &$\mathbf{Y}_i(t)$  &  2 & 1  & 1  & 1\\
 & & &$\bar{{W}}_{i}(t)$ & 2.9 & 0.916  & 0.873  & 0.833\\
 & & &${Y}^{*}_i(t)$ & 2.06  & 0.930 & 0.902 & 0.862\\
 \cline{3-8}
 &  & \multirow{3}{*}{80} & ${Y}_i(t)$  & 2  & 1  &  1 & 1 \\
   &   &  & $\bar{{W}}_{i}(t)$ & 5.4 & 0.933 & 0.866 & 0.866 \\
  &   &  & ${Y}^{*}_i(t)$ &  2.88 & 0.959 & 0.918 & 0.918\\
 \cline{2-8}
& \multirow{6}{*}{3} & \multirow{3}{*}{40} & ${Y}_i(t)$ &  2 & 1 & 1  &  1\\
 & & & $\bar{{W}}_{i}(t)$ & 5.32 & 0.792 & 0.629 & 0.581\\
 & & &${Y}^{*}_i(t)$ & 2.46 & 0.826 & 0.704 & 0.652\\
 \cline{3-8}
 &  & \multirow{3}{*}{80} & ${Y}_i(t)$ & 2 & 1  & 1  & 1  \\
   &   &  & $\bar{{W}}_{i}(t)$ & 8.86 & 0.784 & 0.613 & 0.567 \\
  &   &  & ${Y}^{*}_i(t)$ & 2.76 & 0.833 & 0.707 & 0.666\\
   \hline
  
\end{tabular}
\caption{Clustering results of different methods, where ${Y}_i(t)$ denotes the ORACLE, $\bar{{W}}_{i}(t)$ and ${Y}^{*}_i(t)$ represent the naive method and our proposed method based on the averaged data respectively. }
\label{table:simu_avg}
\end{table}

\begin{table}
\centering
\begin{tabular}{|r|r|r|r|cccc|}
\hline
Distance &$\sigma$ & n & Method & \# clusters & Rand & Jaccard & aRand \\
   \hline
\multirow{12}{*}{Far}& \multirow{6}{*}{2}& \multirow{3}{*}{40} & ${Y}_i(t)$ & 2 & 1  & 1  &  1 \\
 & & &${W}_{i1}(t)$& 4.8 & 0.918 & 0.831 & 0.834\\
 & & &${Y}^{*}_i(t)$ & 2.34  &  0.966  & 0.932  & 0.932\\
 \cline{3-8}
 &    & \multirow{3}{*}{80} & ${Y}_i(t)$ & 2 & 1  & 1  &  1 \\
   &   &  & ${W}_{i1}(t)$ & 7.6 & 0.920 & 0.839 & 0.840\\
  &   &  & ${Y}^{*}_i(t)$ & 2.34  & 0.974  & 0.949  & 0.949\\
\cline{2-8}
& \multirow{6}{*}{3}& \multirow{3}{*}{40} &${Y}_i(t)$ & 2 &  1 & 1  &   1\\
 & & &${W}_{i1}(t)$& 5.76 & 0.836 & 0.681 & 0.669\\
 & & &${Y}^{*}_i(t)$ & 2.32 &  0.894 &  0.802 &  0.787\\
 \cline{3-8}
 &    & \multirow{3}{*}{80} &  ${Y}_i(t)$ & 2 & 1  &  1 & 1  \\
   &   &  & ${W}_{i1}(t)$ & 27.3 & 0.739 & 0.488 & 0.475\\
  &   &  & ${Y}^{*}_i(t)$ &  7.2 & 0.883 & 0.785 & 0.766 \\
   \hline
    \hline
\multirow{12}{*}{Close}& \multirow{6}{*}{2} &  \multirow{3}{*}{40} &${Y}_i(t)$ & 2 & 1  & 1  &  1 \\
 & & &${W}_{i1}(t)$ &  5.44 & 0.732 & 0.536 & 0.461\\
 & & &${Y}^{*}_i(t)$ & 2.78 & 0.772 & 0.612 & 0.542\\
 \cline{3-8}
 &  & \multirow{3}{*}{80} & ${Y}_i(t)$ & 2 &  1 & 1  & 1  \\
   &   &  & ${W}_{i1}(t)$ & 11.44 & 0.705 & 0.521 & 0.409 \\
  &   &  & ${Y}^{*}_i(t)$ & 2.14  & 0.750  & 0.638  & 0.501\\
 \cline{2-8}
& \multirow{6}{*}{3} & \multirow{3}{*}{40} &${Y}_i(t)$ & 2 & 1  & 1  & 1\\
 & & &${W}_{i1}(t)$ & 3.9 &  0.527 & 0.450 &  0.068\\
 & & &${Y}^{*}_i(t)$ & 1.58 & 0.529 & 0.483 & 0.076\\
 \cline{3-8}
 &  & \multirow{3}{*}{80} & ${Y}_i(t)$ & 2 &  1 &  1 & 1  \\
   &   &  & ${W}_{i1}(t)$ &  16.94 & 0.502 & 0.388 & 0.007\\
  &   &  & ${Y}^{*}_i(t)$ &   19.00 &   0.503 &   0.374 &   0.008\\
   \hline
  
\end{tabular}
\caption{Clustering results of different methods, where ${Y}_i(t)$ denotes the ORACLE, ${{W}}_{i1}(t)$ and ${Y}^{*}_i(t)$ represent the naive method and our proposed method based on a single observation respectively. }
\label{table:simu_single}
\end{table}

Tables \ref{table:simu_avg} and  \ref{table:simu_single} present the results with pseudo observations $\left\{{Y}^*_i(t)\right\}_{i=1}^{n}$  simulated from the estimated conditional distribution of ${Y}_i|\bar{{W}}_{i}$ and ${Y}_i|{{W}}_{i1}$, respectively. The ORACLE (${Y}_i(t)$) method performs the best with correct clustering results in all replications. Both Naive and our proposed methods 
exhibit better performance when there is a greater disparity between clusters (Far) or when the variance of measurement error is reduced ($\sigma=2$). 
 However, our proposed methods (${Y}_i^{*}(t)$) consistently demonstrate improved performance than the naive ones ( $\bar{{W}}_{i}(t)$  or ${W}_{i1}(t)$ ) in almost all scenarios. The Naive method tends to yield an excess of subgroups, leading to an unnecessarily larger number of clusters due to measurement errors. In contrast, our proposed method yields an average cluster size that aligns more closely with the actual number of clusters, providing a more accurate representation of the underlying data structure. The observation is also supported by the fact our method gives higher values of Rand, Jaccard, and aRand indexes than the naive ones. However, for one case in Table \ref{table:simu_single} with  $n=80$, $\sigma=3$ and $Close$, our proposed method does not yield improved results in this scenario. This is primarily due to the limitation of the naive method, which fails to generate meaningful clustering results suitable as initial values for our proposed approach.

\begin{figure}
\centering
\includegraphics[scale=0.45]{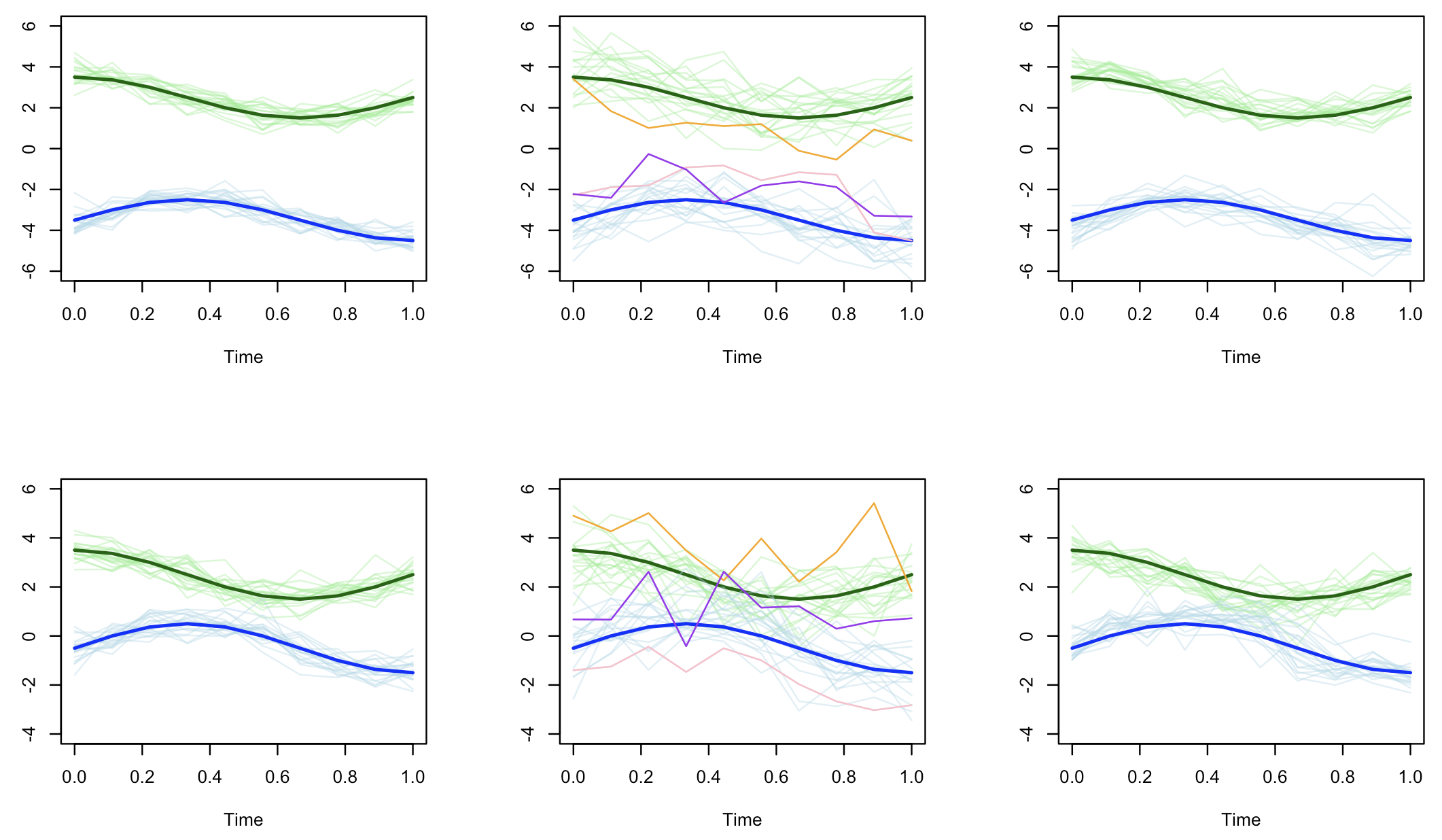}
\caption{Simulation results based on one replication for the case $n=40, \sigma=2$. The top panel corresponds to "far" and the bottom panel corresponds to "close". Different colors represent different groups. Two bold curves, dark green and dark blue, represent the true mean functions of the two groups. The left panels plot the true functional responses ${Y}_i(t)$. The middle panels give the observed average data $\bar{{W}}_{i}(t)$, with five groups identified using the naive method for both cases. The right panels plot the simulated data ${Y}_i^{*}(t)$ from our proposed method, where it correctly identifies two groups. } \label{fig:sim}
\end{figure}

In Figure \ref{fig:sim}, we present the simulation results for a single replication for data generated under "far" (top panel) and "close" (bottom panel) with  $n=40$, $\sigma=2$.  It compares the true functional data (left panel),  the observed average functional data (middle panel) from three repeated measures, and the simulated functional data (left panel) using our proposed method. It also presents the clustering results using the naive method in the middle panel and the proposed method in the left panel. Figure \ref{fig:sim} shows that our proposed method effectively removes the extra variability due to measurement errors. As a result, it correctly identifies two functional subgroups and absorbs the three singleton subgroups produced by the naive method into the correct subgroups.  

\section{REAL DATA APPLICATION}

In this section, we apply our proposed method to a data set from a children's obesity study conducted by Dr. Mark Benden and colleagues from 2012 to 2014 in the College Station Independent School District of Texas. In this dataset, daily energy expenditure, $Y(t)$, is defined as the total number of calories or energy used by the body to perform everyday bodily functions. However, the true values of $Y(t)$ are not directly observable. Instead, a surrogate measure for daily energy expenditure, $W(t)$, was collected per minute using the Sense Wear Armband ${ }^{\circledR}$ (BodyMedia, Pittsburgh, PA) for students who wore accelerometers while in school for one week (five days) at baseline. 

We follow the data preprocessing procedures outlined in \citep{Zhangpl2023} for handling missing data and data centralization. Specifically, missing values in energy expenditure measurements for each individual are imputed using a cubic splines regression model. Furthermore, to mitigate the potential influence of enrollment time disparities across participating schools, the measured energy expenditure data was centered by subtracting the daily averaged energy expenditure from each individual curve.

\begin{figure}
\centering
\includegraphics[scale=0.35]{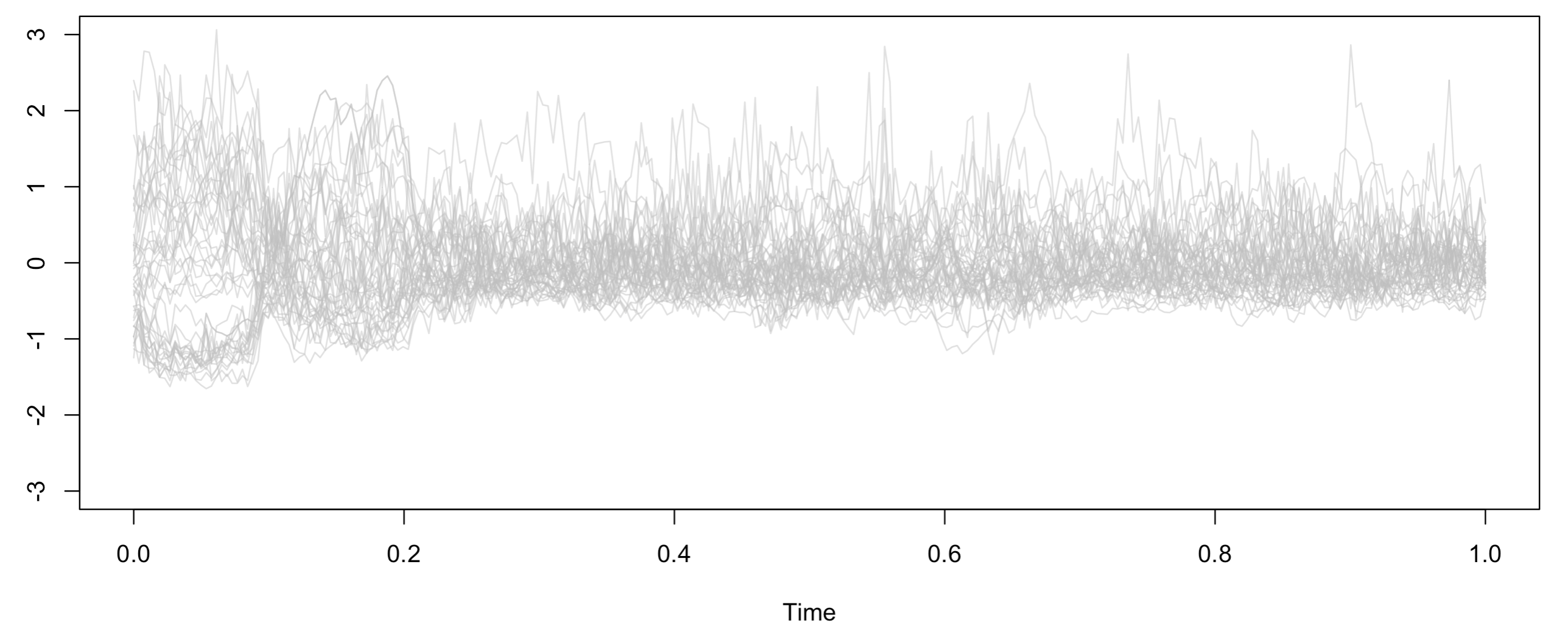}
\caption{Plots of observed average energy expenditure $\bar{{W}}_{i}(t)$ versus time for all individuals.}\label{fig:realdata_avg}
\end{figure}

In this study, we focus on the dataset consisting of 45 third-grade students from a randomly selected school. Due to limited data availability, some students only have three days of energy expenditure measurements. To ensure consistency, we select three days' data as the repeated measurements for $Y(t)$. Figure \ref{fig:realdata_avg} illustrates the average energy expenditure measurements over this three-day period, denoted as $\bar{{W}}_{i}(t)=\sum_{j=1}^{3}W_{ij}(t)/3$, plotted against time for all individuals.

It can be seen that in the latter half of the time frame, students' energy expenditure exhibits relatively stable patterns. In light of this observation, our subsequent analyses will focus specifically on the initial half of the time frame. To reduce the variability, we smooth the daily energy expenditure curve by averaging measurements at five-minute intervals, resulting in 27 data points for each curve. Additionally, the time interval is scaled to $[0,1]$. 

Clustering analysis is first applied to the average energy expenditure measurements $\bar{W}_{i}(t)$. Then, the proposed method is applied following the algorithm outlined in Section 3.1, with pseudo observations $\left \{{Y}^*_i(t)\right \}_{i=1}^{n}$ simulated from the estimated conditional distribution of ${Y}_i|\bar{{W}}_{i}$. In both clustering methods, the linear spline is used, and the number of internal knots is set to $10$ to provide an adequate approximation of the functional data.

\begin{figure}
\centering
\includegraphics[scale=0.35]{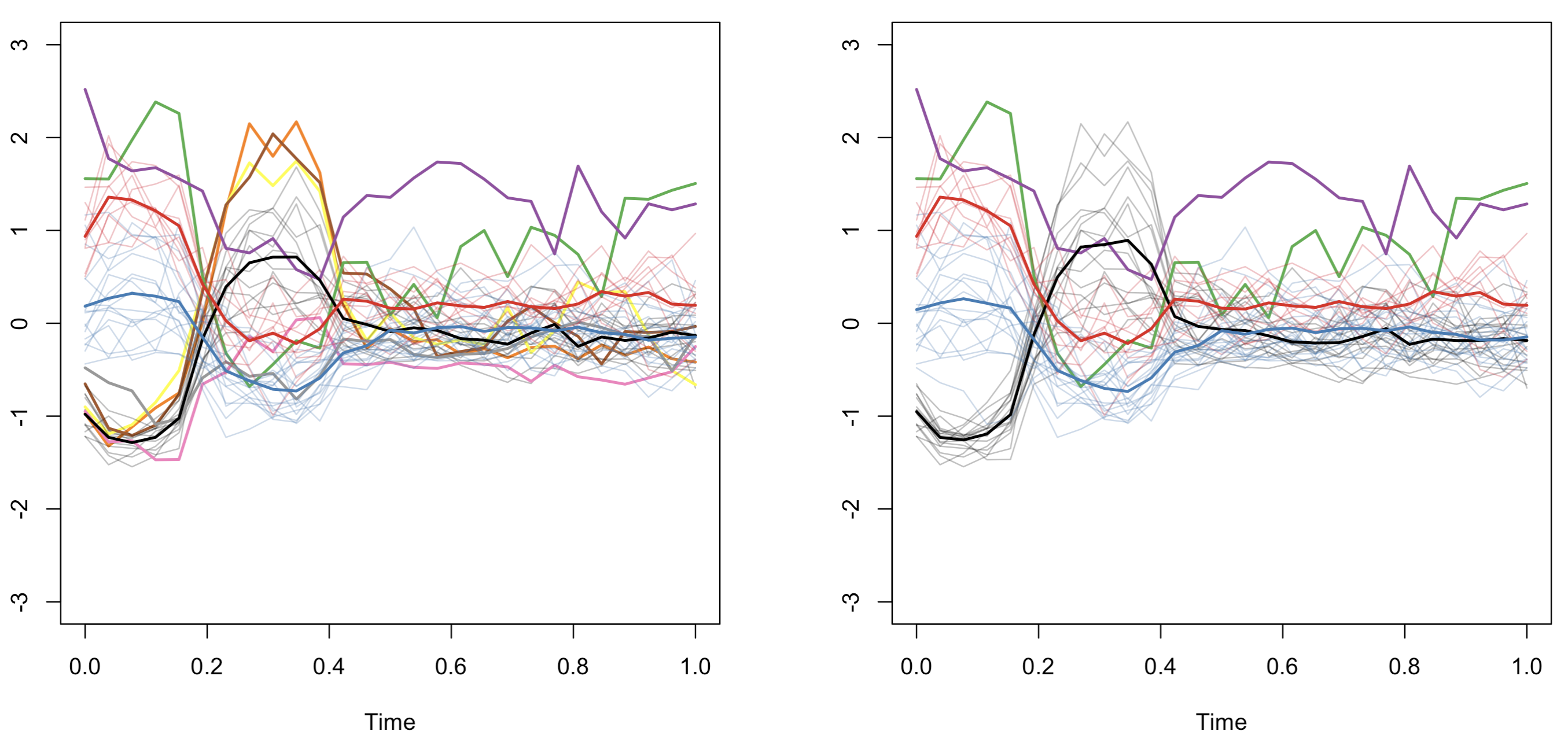}
\caption{Initial clustering results from the average data $\bar{{W}}_{i}(t)$(left panel) and the clustering results of the proposed method (right panel). Each distinct group is represented by a unique color. The bold curves are the group average curves. } \label{fig:realdata}
\end{figure}

The left panel of Figure \ref{fig:realdata} illustrates the clustering results on the average data $\bar{{W}}_{i}(t)$ over three repeated measurements. Different groups are distinguished by distinct colors, and the bold curves indicate the group averages. A total of ten groups are identified, with the red, black, and blue clusters being three big groups, while the remaining groups each consist of a single individual and stand as singletons.

The right panel presents the clustering results generated by our proposed method, using initial clusters identified from $\bar{{W}}_{i}(t)$. It is evident that our approach incorporates measurement errors and yields a more concise classification, revealing only five groups. Most individuals maintain their original group assignments from the initial clustering, with the exception of five singletons (yellow, orange, brown, pink, and gray) which are absorbed into larger groups.  This integration of the groups makes sense, for example, as can be seen from the figure, three singletons (yellow, orange, brown), have a similar pattern as those individuals in the black group.

\section{DISCUSSION}
We studied clustering functional data with measurement errors. Most existing approaches in the literature perform clustering analysis of functional data when there is no measurement error. In our proposed methods, we considered the function-valued data, $W(t)$, to be a surrogate or observed measure of a latent function-valued covariate, $Y(t)$, prone to heteroscedastic classical measurement error. We used repeated measures of the surrogate to estimate the distribution of the functional measurement errors, and a simulation-based approach was used to simulate the error-corrected function-valued data. Through simulations, we compared the finite sample performance of our proposed method to those obtained under the Oracle method, $Y(t)$, and the naive uncorrected approach, $W(t)$. Our results illustrated the importance of correcting for measurement error when there is a considerable amount of measurement error associated with the function-valued covariate.  We successfully applied our methods to a childhood obesity study and performed a clustering analysis based on wearable device-based measures of physical activity for the students included in this study. Our proposed method yielded more dependable clustering results, enhancing our ability to discern meaningful patterns in children's physical activity. These results not only contribute to a better understanding of children's physical activity behaviors but also offer valuable insights for further investigations into the relationship between physical activity and obesity.



\section*{ACKNOWLEDGEMENTS}

{\it Conflict of Interest}: None declared.
This research was supported by National Institute of Diabetes and Digestive and Kidney Diseases Award 1R01DK132385-01. 

\bibliography{reference}

\clearpage

\end{document}